\documentclass[%
 reprint,
 amsmath,amssymb,
 aps,
]{revtex4-2}

\usepackage{graphicx}
\usepackage{dcolumn}
\usepackage{bm}
\graphicspath{{figures/}}
\usepackage{epstopdf}
\usepackage[separate-uncertainty=true]{siunitx} 
\DeclareSIUnit\gauss{Gauss}
\DeclareSIUnit\mbar{\milli\bar}
\usepackage{algorithm}
\usepackage{amsmath}
\usepackage{amsfonts}
\usepackage{array}
\usepackage{dsfont}
\usepackage{booktabs} 
\usepackage{tabularx}
\usepackage{multirow}
\usepackage{adjustbox}
\usepackage{ragged2e} 
\usepackage[version=3]{mhchem} 
\usepackage{physics}
\usepackage{algpseudocode}
\usepackage{hyperref}   
\usepackage[noabbrev]{cleveref} 
\graphicspath{{figures/}}
\usepackage{dcolumn}
\usepackage{bm}
\usepackage{graphicx} 
\usepackage{pgfornament}

\usepackage{pgfplots}

\usepgfplotslibrary{groupplots}

\pgfplotsset{compat=newest}

\usepgfplotslibrary{fillbetween}

\pgfplotsset{select coords between index/.style 2 args={
		x filter/.code={
			\ifnum\coordindex<#1\fi
			\ifnum\coordindex>#2\fi
		}
}}

\pgfplotsset{every axis/.append style={
		label style={font=\small},
		tick label style={font=\small},
		mark size={1.5pt}  
}}

\usepackage{braket}
\usepackage{amsmath}
\usepackage[thinc]{esdiff}
\usepackage{graphicx, color}
\usepackage{tikz}
\definecolor{markgrijs}{RGB}{92,92,92}
\definecolor{markrood}{RGB}{139,0,0}
\definecolor{markblauw}{RGB}{132,189,195}
\definecolor{markdiepblauw}{RGB}{43,92,220}
\definecolor{markgoud}{RGB}{202,171,49}
\definecolor{markviesblauw}{RGB}{138,162,158}
\definecolor{marklichtgrijs2}{RGB}{200,200,200}
\definecolor{rainbow1of8}{RGB}{86,152,198}
\definecolor{rainbow2of8}{RGB}{255,158,74}
\definecolor{rainbow3of8}{RGB}{96,183,96}
\definecolor{rainbow4of8}{RGB}{224,92,93}
\definecolor{rainbow5of8}{RGB}{174,140,205}
\definecolor{rainbow6of8}{RGB}{168,128,119}
\definecolor{rainbow7of8}{RGB}{223,152,209}
\definecolor{rainbow8of8}{RGB}{158,158,158}
\definecolor{rainbow9of8}{RGB}{204,205,89}
\definecolor{rainbow10of8}{RGB}{80,206,218}

\begin{document}

\preprint{APS/123-QED}

\title{On-the-Spot Loading of Single-Atom Traps}
\author{Mark IJspeert}
 \affiliation{Clarendon Laboratory, University of Oxford, Parks Road, Oxford, OX1 3PU, UK}

\author{Naomi Holland}%
\affiliation{Clarendon Laboratory, University of Oxford, Parks Road, Oxford, OX1 3PU, UK}%

\author{Benjamin Yuen}
\affiliation{Clarendon Laboratory, University of Oxford, Parks Road, Oxford, OX1 3PU, UK}

\author{Axel Kuhn}
\email{axel.kuhn@physics.ox.ac.uk}
\affiliation{Clarendon Laboratory, University of Oxford, Parks Road, Oxford, OX1 3PU, UK}%

\date{\today}

\begin{abstract}
	Reconfigurable arrays of trapped single atoms are an excellent platform for the simulation of many-body physics and the realisation of high-fidelity quantum gates. The confinement of atoms is often achieved with focussed laser beams acting as optical dipole-force traps that allow for both static and dynamic positioning of atoms.	In these traps, light-assisted collisions---enhancing the two-atom loss rate---ensure that single atom occupation of traps can be realised. However, the time-averaged probability of trapping a single atom is limited to $0.5$ when loading directly from a surrounding cloud of laser-cooled atoms, preventing deterministic filling of large arrays. In this work, we demonstrate that increasing the depth of a static, optical dipole trap enables the transition from fast loading on a timescale of $2.1\,$s to an extended trap lifetime of $7.9\,$s. This method demonstrates an achievable filling ratio of $\SI{79\pm2}{\percent}$ without the need of rearranging atoms to fill vacant traps.
\end{abstract}

\maketitle

\section{\label{intro:level1}Introduction}
Single trapped atoms at temperatures near absolute zero constitute ideal information carriers in quantum computing and simulation. Recent advances in this field underpin the utility of using reconfigurable, optical tweezers to trap cold neutral atoms. This approach has enabled remarkable achievements in the context of quantum information science and many-body physics \cite{Labuhn2016, Endres2016, Bernien2017, Lienhard2018, Barredo2018}. Apart from being highly scalable \cite{Manetsch2024}, neutral atoms in reconfigurable microtraps exhibit long coherence times \cite{Norcia2019,Wu2019,Young2020} and allow for arbitrary qubit connectivity \cite{Beugnon2007,Bluvstein2022}. They furthermore enable fully programmable single-qubit rotations \cite{Ma2023} and are shown to yield a high readout and two-qubit gate fidelity \cite{Evered2023}. The combination of these characteristics fostered the development of a logical quantum processor based on reconfigurable atom arrays \cite{Bluvstein2023}. In the context of quantum repeaters and networks, arrays of individually controlled neutral atoms coupled to optical cavities are equally promising for generating efficient remote entanglement over a mesh of quantum nodes \cite{Wilk2007, rempe2012, Covey2023}. 

Single-atom optical dipole traps have been implemented using a large range of different techniques with varying degrees of configurability: acousto-optic deflectors \cite{Endres2016, Bernien2017}, micro-lens arrays \cite{Pause2024}, standing wave dipole traps \cite{Alt2003, Gallego2018}, digital mirror devices (DMD) \cite{Stuart2018} and liquid crystal spatial light modulators (SLM) \cite{Bergamini2004, Kim2016}. The use of a DMD or liquid crystal SLM for holographic beam shaping extends the range of trapping geometries beyond periodic lattices to arbitrary configurations in up to three dimensions \cite{Barredo2018}.

With any of these trapping techniques, achieving consistent single-atom occupation across multiple trapping sites constitutes a challenging task, due to the fact that atoms enter the traps stochastically \cite{Grimm2000}. To be precise, in sufficiently tight dipole traps---those with a beam waist of around \SI{1}{\micro\meter}---single-atom loading is accomplished by virtue of light-assisted collisions, which cause a collisional blockade \cite{Schlosser2002}. This regime is characterised with a strong two-atom loss rate due to inelastic collisions, resulting in a negligible probability of finding a doubly occupied trap. When loading atoms directly from a magneto-optical trap (MOT) or optical molasses however, the time-averaged, single-atom occupation probability is limited to $0.5$. This implies that the likelihood of establishing fully occupied arrays of traps will decrease exponentially as the system size increases. The strategies to overcome this problem are to either increase the single-atom loading efficiency \cite{Fung2014} or to relocate ancillary atoms to vacant positions \cite{Vala2005}. Whilst this is typically accomplished with acousto-optic beam deflectors \cite{Barredo2018}, atom relocation has also been achieved through the use of dynamic holograms \cite{Kim2016,Kim2019}. When atoms are transported by means of dynamic holograms however, phase changes of a liquid crystal SLM or re-settling oscillations of DMD micro-mirrors give rise to uncontrollable fluctuations in the trapping potential, causing a dramatic increase in the atom loss rate \cite{Stuart2018,McGloin2003}.

In this work, we investigate the effect of changing the depth of a static, optical dipole trap on the atom loading and loss rates and propose a feedback mechanism that enhances the trap filling ratio greatly beyond what is maximally achievable in the collisional blockade regime. Apart from requiring significantly less optical power with respect to the conventional approach of relocating atoms, this method reduces the need for atom rearrangement to achieve consistent filling of dipole trap arrays. 

\section{\label{theory:level1}Theory}
Fundamental to trapping atoms in optical potentials is the dipole interaction between matter and light. The interaction Hamiltonian can be written as
\begin{equation}
H_I = -\vec{d}\cdot \vec{\mathcal{E}}(t)
\end{equation}
in terms of the electric dipole operator $\vec{d}$ and an applied external electric field $\vec{\mathcal{E}}(t)$. For a two-level system considered in the dressed-state picture, this interaction normally leads to an energy shift of the ground state by $\Delta E={\hbar \vert\Omega\vert^2}/{4\Delta}$, where $\Omega$ is the Rabi frequency of the laser driving the transition, detuned by $\Delta=\omega-\omega_0$ from the atomic transition frequency $\omega_0$. Considering the entire manifold of states in Rubidium, the dipole interaction must be considered as a perturbation to the atomic states in order to calculate the ac Stark shifts that arise from it. Depending on the sign of the perturbation, these energy shifts give rise to either attractive or repulsive optical potentials towards the focus for atoms in their ground state. It follows from the conservation of parity that the first-order perturbation to the energy levels vanishes for alkali atoms. Therefore, the Stark shift $\Delta E_{J,m_J}$ of a state $|J,m_J\rangle$ with energy $E_{J,m_J}$ is given by the second-order perturbation
\begin{equation}
	\Delta E_{J,m_J} = \sum_{J', m_J'} \frac{\langle J,m_J | H_I | J',m_J' \rangle \langle J',m_J' | H_I | J,m_J \rangle}{E_{J,m_J} - E_{J',m_J'}}
\end{equation}
which includes the sum over all intermediate states allowed by electric-dipole transition selection rules. The energy shifts can be rewritten in terms of the scalar and tensor polarisabilities ($\alpha_0(\omega)$ and $\alpha_2(\omega)$) using the Wigner-Eckart theorem as
\begin{equation}
	\Delta E_{J,m_J} = -\frac{1}{4}\left(\alpha_0(\omega)+\alpha_2(\omega) \frac{3m_J^2-J(J+1)}{J(2J-1)}\right) \mathcal{E}_0^2
\end{equation}
for linearly polarised light of frequency $\omega$ and field amplitude $\mathcal{E}_0$, detuned by at least several linewidths from resonance. If the detuning is large compared to the excited-state hyperfine splitting, the tensor component of the ground state polarisability vanishes and the scalar polarisability can be approximated as
\begin{equation}
	\alpha_0(\omega)\approx\sum_{J'}\frac{2\omega_{J'J} |\langle J=1/2 \Vert \vec{d}\, \Vert J' \rangle|^2}{3\hbar (\omega_{J'J}^2 - \omega^2)}
\end{equation}
written in terms of the reduced dipole matrix elements for the transitions with frequencies $\omega_{J'J}=(E_{J'}-E_J)/\hbar$. For \ce{^{87}Rb} atoms, a red-detuned beam at $1064\,$nm renders the denominators in the sum positive and hence the overall negative shift of the ground state energy gives rise to an attractive trapping potential. Since the resulting dipole-force traps are conservative, loading requires atoms to be cooled in a magneto-optical trap (MOT). Once loaded into the trap, there are two mechanisms by which atoms are lost. The first is via a collision with a background atom that imparts sufficient kinetic energy for the atom's energy to exceed the depth of the trap. The second loss mechanism comes into play when a light-assisted collision between two trapped atoms causes both of the atoms involved to be lost from the trap. For a sufficiently small trapping volume, the two-body loss rate dominates and the time-averaged, single atom occupation is limited to $0.5$.
\section{Experimental apparatus}
\begin{figure*}
	\includegraphics[width=\textwidth]{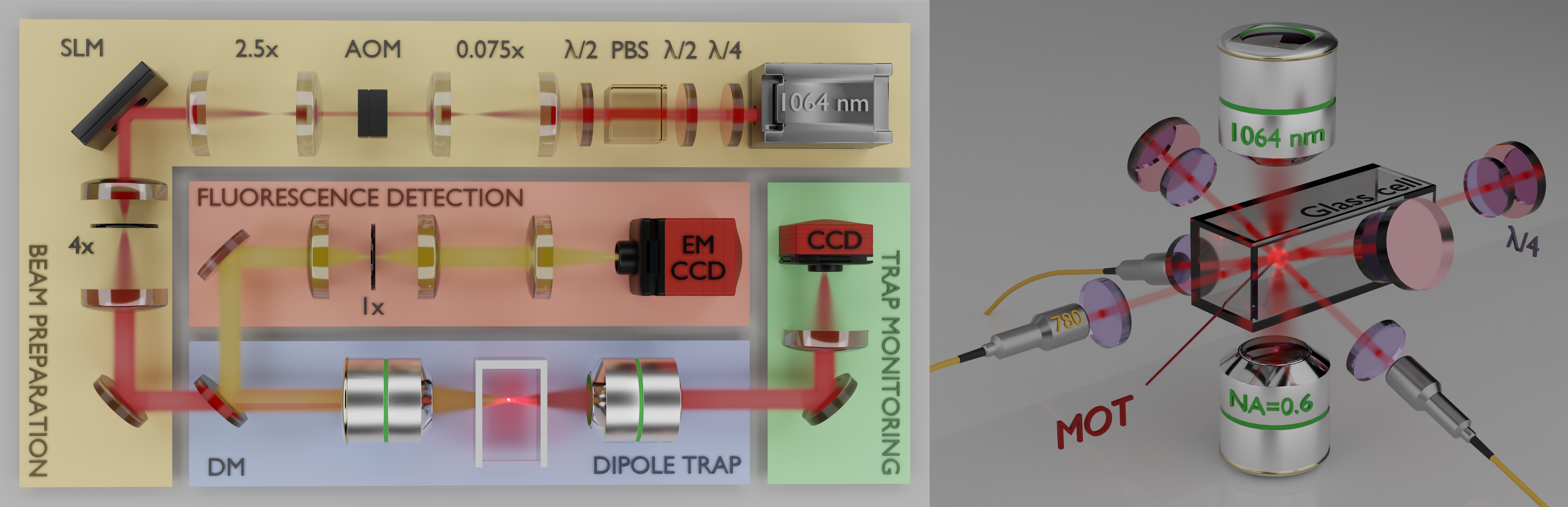}
	\caption{(Left) Optical setup for the generation of $1064\,$nm dipole traps and the detection of fluorescence. A spatial light modulator (SLM) generates a $2\times4$ tweezer array inside the glass cell, monitored via a CCD. A dichroic mirror (DM) redirects the fluorescence into a detection path, where atoms are imaged using an EMCCD camera. Numbers indicate the beam magnification factors. (Right) Optical setup for the generation of the MOT inside a glass cell using $3$ retro-reflected beams. A pair of high-NA lenses focus and image the dipole trapping light.}
	\label{fig:exptSetup}
\end{figure*}

In our experiment, \ce{^{87}Rb} atoms are cooled in a MOT on the \ce{D2} transition ($\ce{^2S_{1/2}} \, \leftrightarrow \, \ce{^2P_{3/2}}$ at \SI{780.246}{\nm}) and eventually loaded into optical dipole-force traps at a wavelength of \SI{1064}{\nm}. The MOT is formed using $3$ retro-reflected, circularly polarised beams (\autoref{fig:exptSetup}), each consisting of overlapping cooling and repump beams with a $1/e^2$ beam diameter of \SI{5}{\milli\meter} and optical powers of \SI{1.5}{\mW} and \SI{0.5}{\mW}, respectively. The cooling light is red-detuned from the $F=2 \, \leftrightarrow \, F'=3$ transition and is frequency modulated between $-1.7\Gamma$ and $-2.7\Gamma$ from resonance at a modulation frequency of \SI{100}{\kHz}. This is done to eliminate the impact of interference fringes and thus achieve a homogeneously filled cloud of atoms from which to load the dipole traps. The re-pumping light is tuned to the $F=1 \, \leftrightarrow \, F'=2$ transition. The MOT is positioned within a glass cell vacuum chamber (ColdQuanta), with cell walls AR coated for both \SI{780}{\nm} and \SI{1064}{\nm} (\SI{< 0.2}{\percent} for $\SI{650}{\nm} \leq \lambda \leq \SI{1100}{\nm}$ and angle of incidence between \SI{\pm 10}{\degree}, \SI{< 0.5}{\percent} for $\SI{760}{\nm} \leq \lambda \leq \SI{860}{\nm}$ and angle of incidence between \SI{\pm 45}{\degree}). The magnetic field is generated using a pair of anti-Helmholtz coils which are oriented along the long axis of the glass cell and establish a field gradient of \SI{13}{\gauss\per\cm} at the quadrupole centre. The laser that generates the dipole traps is a CW fibre laser (Quantel EYLSA), which outputs single mode, narrow linewidth (\SI{<700}{\kHz}) light at \SI{1064}{\nm}, featuring an $M^2$ beam quality factor of \num{1.6}. The trapping light is first sent through an acousto-optic modulator (AOM) before a computer-generated phase pattern is imposed via a $512\times512$ Meadowlark liquid crystal SLM, positioned in a Fourier plane of the light path. This can be viewed as an artificial hologram that gives rise to an intensity pattern in the image plane. This pattern, which contains all dipole traps, can be identified as the Fourier transformed light field that passes the SLM. By using an iterative approach based on feedback from the measured intensity of the generated trapping sites in combination with the mixed-region amplitude freedom (MRAF) algorithm \cite{Pasienski2008} to generate arbitrary phase patterns, we construct a uniform $2\times4$ tweezer grid in the focal plane with $1/\mathrm{e}^2$ trap diameters of \SI{2.2}{\micro\meter}, allowing for parallel data acquisition from multiple traps. A pair of achromatic, compound lens systems ($\mathrm{NA} = 0.6$) focus and re-collimate the trapping light, such that traps can be monitored using a CCD camera. The numerical aperture of \num{0.6} equates to a \SI{10}{\percent} collection efficiency of the fluorescence light, which is directed towards an EMCCD camera (Princeton Instruments ProEM-HS: 512BX3) via a dichroic mirror (Thorlabs DMLP950L) with a cut-off wavelength of \SI{950}{\nm} to block any trapping light from reaching the camera. In addition, a \SI{780}{\nm} bandpass filter (Semrok \num{780}/\SI{12}{\nm} BrightLine single-band) is placed directly in front of the EMCCD camera and eliminates any remaining light at \SI{1064}{\nm}.
\section{Trap lifetimes}
\begin{figure}[h]
	\centering
	\begin{tikzpicture}
	
	\begin{groupplot}[group style={group size=3 by 2, horizontal sep=0.5cm, vertical sep=1.0cm},xmin=0,ymin=0,height=3.7cm,width=3.7cm,no markers]
	
	\nextgroupplot
	[
	ylabel={Counts (MHz)},
	xlabel = {Time (s)},
	xmin=25, xmax=45,
	ymin=-100/1000, ymax=500/1000,
	xtick={25,30,35,40,45},
	xticklabels={0,5,10,15,20},
	xmajorgrids=true,
	ymajorgrids=true,
	grid style=dashed,
	y label style={at={(axis description cs:-0.17,.5)},anchor=south},
	y tick label style={rotate=90},
	x label style={at={(axis description cs:0.5,-0.17)},anchor=north},
	]
	
	\addplot[
	color=rainbow4of8
	] table [col sep = comma, x index = 0, y expr=\thisrowno{1}*1.3062923961932331*9/1000/1000/0.033]{TRACE21b.csv};
	\node[anchor=west] at (rel axis cs:0.7,0.85) {\small (a)};
	
	\nextgroupplot
	[
	xlabel = {Counts (MHz)},
	xmin=-100/1000, xmax=700/1000,
	yticklabels={,,},
	xmajorgrids=true,
	grid style=dashed,
	scaled y ticks = false,
	x label style={at={(axis description cs:0.5,-0.17)},anchor=north},
	]
	
	\addplot[
	ybar,
	bar width=1.1pt,
	bar shift=0.55pt,
	fill=markgrijs,
	draw=markgrijs,
	mark = *,
	color=markgrijs,
	opacity=0.5
	] table [col sep = comma, x expr=\thisrowno{0}*1.3062923961932331*9/1000/1000/0.033, y index = 1]{FLUORhist21.csv};
	
	\addplot[
	color=rainbow4of8,
	style={thick}
	] table [col sep = comma, x expr=\thisrowno{0}*1.3062923961932331*9/1000/1000/0.033, y index = 1]{FLUORFIThist21.csv};
	
	\node[anchor=north] (source) at (axis cs:794*1.3062923961932331*9/1000/1000/0.033,0.0042){};
	\draw[-latex, color=rainbow4of8](source)--++(0cm, -0.5cm);
	\node[anchor=west] at (rel axis cs:0.7,0.85) {\small (b)};
	
	\nextgroupplot
	[
	xlabel = {Lifetime (s)},
	xmin=-0, xmax=10,
	yticklabels={,,},
	xmajorgrids=true,
	grid style=dashed,
	scaled y ticks = false,
	x label style={at={(axis description cs:0.5,-0.17)},anchor=north},
	]
	
	\addplot[
	ybar,
	bar width=1.1pt,
	bar shift=0.55pt,
	fill=markgrijs,
	draw=markgrijs,
	mark = *,
	color=markgrijs,
	opacity=0.5
	] table [col sep = comma, x expr=\thisrowno{0}/30, y index = 1]{LIFEhist21.csv};
	
	\addplot[
	color=rainbow4of8,
	style={thick}
	] table [col sep = comma, x expr=\thisrowno{0}/30, y index = 1]{LIFEFIThist21.csv};
	
	\node[anchor=west, color=rainbow4of8] at (axis cs: 2, 4.5/1000) {\small ${\tau}=5.6\,$s};
	\node[anchor=west] at (rel axis cs:0.7,0.85) {\small (c)};

	\nextgroupplot
	[
	xlabel={Time (s)},
	ylabel={Counts (a.u.)},
	xmin=45, xmax=65,
	ymin=0, ymax=150000,
	yticklabel style={
		/pgf/number format/fixed,
		/pgf/number format/precision=5,
		/pgf/number format/fixed zerofill
	},
	scaled y ticks=false,
	yticklabels={,,},
	xtick={45,50,55,60,65},
	xticklabels={0,5,10,15,20},
	ymajorgrids=true,
	xmajorgrids=true,
	grid style=dashed,
	x label style={at={(axis description cs:0.5,-0.17)},anchor=north},
	y label style={at={(axis description cs:-0.17,.5)},anchor=south}
	]
	
	\addplot[
	mark size=1pt,
	name path=dtdatapoints,
	color=rainbow1of8,
	/pgfplots/error bars/.cd,
	y dir = both,
	y explicit,
	] table [col sep = comma, x index = 0, y index = 2]{tracesofdark_1.csv};

	\node[anchor= center, font=\small, color=rainbow1of8, align=center] at (axis cs: 55, 20000) {$\Delta t=\SI{0.5}{\s}$};
	\node[anchor=west] at (rel axis cs:0.7,0.85) {\small (d)};

	\nextgroupplot
	[
	xlabel={Time (s)},
	xmin=45, xmax=65,
	ymin=0, ymax=150000,
	yticklabel style={
		/pgf/number format/fixed,
		/pgf/number format/precision=5,
		/pgf/number format/fixed zerofill
	},
	scaled y ticks=false,
	xtick={45,50,55,60,65},
	xticklabels={0,5,10,15,20},
	yticklabels={,,},
	ymajorgrids=true,
	xmajorgrids=true,
	grid style=dashed,
	x label style={at={(axis description cs:0.5,-0.17)},anchor=north},
	]
	
	\addplot[
	mark size=1pt,
	name path=dtdatapoints,
	color=rainbow2of8,
	/pgfplots/error bars/.cd,
	y dir = both,
	y explicit,
	] table [col sep = comma, x index = 0, y index = 2]{tracesofdark_20.csv};

	\node[anchor= center, font=\small, color=rainbow2of8, align=center] at (axis cs: 55, 20000) {$\Delta t=\SI{1}{\s}$};
	\node[anchor=west] at (rel axis cs:0.7,0.85) {\small (e)};
	
	\nextgroupplot
	[
	xlabel = {Time (s)},
	xmin=0, xmax=20,
	yticklabel style={
		/pgf/number format/fixed,
		/pgf/number format/precision=5,
		/pgf/number format/fixed zerofill
	},
	scaled y ticks=false,
	yticklabels={,,},
	ymin=0, ymax=150000,
	legend pos=north west,
	ymajorgrids=true,
	xmajorgrids=true,
	grid style=dashed,
	x label style={at={(axis description cs:0.5,-0.17)},anchor=north},
	]

	\addplot[
	name path=originaldt,
	color=rainbow3of8,
	/pgfplots/error bars/.cd,
	y dir = both,
	y explicit,
	] table [col sep = comma, x index = 0, y index = 2]{tracesofdark_39.csv};
	
	\node[anchor= center, font=\small, color=rainbow3of8, align=center] at (axis cs: 10, 20000) {$\Delta t=\SI{2}{\s}$};
	\node[anchor=west] at (rel axis cs:0.7,0.85) {\small (f)};
	
	\end{groupplot}
	\end{tikzpicture}
	\caption[An example of a floating figure]{(a) Background-corrected fluorescence trace from a dipole trap measured in EMCCD photon counts. Background counts amount to about \SI{0.5}{\mega\hertz} of the signal before subtraction. (b) Histogram of detected counts fitted with a bimodal distribution showing high-fidelity detection of single atom occupation. (c) Exponential fit to the trap occupation-time histogram with a \SI{1}{\second} time bin resolution. (d-f) To measure the survival probability of an atom in the dark, we interrupt the cooling light for a period of $\Delta_t$. Thereafter, the light is switched back on to check for the survival of the atom. This sequence is repeated until the atom is lost. The fluorescence traces are shown for three different values of $\Delta t$.} 
	\label{Fluorfigure1}
\end{figure}
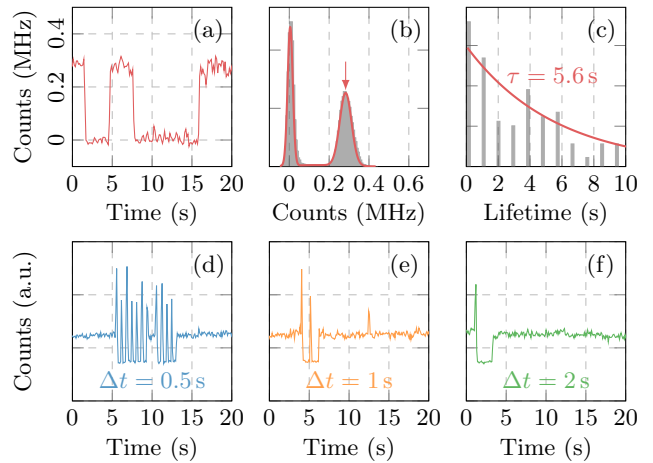

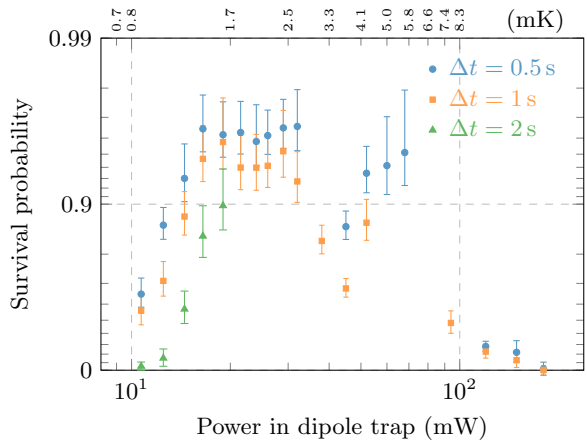
\begin{figure}[h]
	\centering
	\begin{tikzpicture}
	
	\begin{groupplot}[group style={group size=1 by 1,horizontal sep=0.5cm, vertical sep=1.0cm},xmin=0,ymin=0,height=6cm,width=8cm]	
	
	\nextgroupplot
	[
	xmode=log,
	xlabel={Power in dipole trap (mW)},
	ylabel={Survival probability},
	ymin=0, ymax=2,
	ytick={0,1,2},
	minor ytick={0        ,  0.04575749,  0.09691001,  0.15490196,  0.22184875,
		0.30103   ,  0.39794001,  0.52287875,  0.69897   ,  1.        ,
		1.04575749,  1.09691001,  1.15490196,  1.22184875,  1.30103   ,
		1.39794001,  1.52287875,  1.69897   ,  2.        }, 
	yticklabels={0, 0.9, 0.99},
	legend pos=north east,
	legend cell align={left},
	legend style={draw=none,fill=none,text opacity=1},
	ymajorgrids=true,
	xmajorgrids=true,
	grid style=dashed,
	clip=false
	]
	
	\addplot[
	only marks,
	mark = *,
	mark size=1.2pt,
	color=rainbow1of8,
	/pgfplots/error bars/.cd,
	y dir = both,
	y explicit,
	] table [col sep = comma, x index = 0, y expr=\thisrowno{1}, y error minus expr=\thisrowno{1}-\thisrowno{2}, y error plus expr=\thisrowno{3}-\thisrowno{1}]{shortwindowc.csv};
	
	\addplot[
	only marks,
	mark = square*,
	mark size=1pt,
	color=rainbow2of8,
	/pgfplots/error bars/.cd,
	y dir = both,
	y explicit,
	] table [col sep = comma, x index = 0, y expr=\thisrowno{1}, y error minus expr=\thisrowno{1}-\thisrowno{2}, y error plus expr=\thisrowno{3}-\thisrowno{1}]{mediumwindowc.csv};
	
	\addplot[
	only marks,
	mark = triangle*,
	mark size=1.5pt,
	color=rainbow3of8,
	/pgfplots/error bars/.cd,
	y dir = both,
	y explicit,
	] table [col sep = comma, x index = 0, y expr=\thisrowno{1}, y error minus expr=\thisrowno{1}-\thisrowno{2}, y error plus expr=\thisrowno{3}-\thisrowno{1}]{longwindowc.csv};
	
	\addlegendentry{\small \hspace{0.3em}\color{rainbow1of8}$\Delta t=\SI{0.5}{\s}$}
	\addlegendentry{\small \hspace{0.3em}\color{rainbow2of8}$\Delta t=\SI{1}{\s}$}
	\addlegendentry{\small \hspace{0.3em}\color{rainbow3of8}$\Delta t=\SI{2}{\s}$}
	
	\node[anchor=center, color=black, rotate=90] at (axis cs: 9,2.11) {\tiny 0.7};
	\node[anchor=center, color=black, rotate=90] at (axis cs: 10,2.11) {\tiny 0.8};
	\node[anchor=center, color=black, rotate=90] at (axis cs: 20,2.11) {\tiny 1.7};
	\node[anchor=center, color=black, rotate=90] at (axis cs: 30,2.11) {\tiny 2.5};
	\node[anchor=center, color=black, rotate=90] at (axis cs: 40,2.11) {\tiny 3.3};
	\node[anchor=center, color=black, rotate=90] at (axis cs: 50,2.11) {\tiny 4.1};
	\node[anchor=center, color=black, rotate=90] at (axis cs: 60,2.11) {\tiny 5.0};
	\node[anchor=center, color=black, rotate=90] at (axis cs: 70,2.11) {\tiny 5.8};
	\node[anchor=center, color=black, rotate=90] at (axis cs: 80,2.11) {\tiny 6.6};
	\node[anchor=center, color=black, rotate=90] at (axis cs: 90,2.11) {\tiny 7.4};
	\node[anchor=center, color=black, rotate=90] at (axis cs: 100,2.11) {\tiny 8.3};
	
	\node[anchor=center, color=black] at (axis description cs: 0.9,1.05) {(mK)};
	
	\end{groupplot}

\end{tikzpicture}
\caption[An example of a floating figure]{Measured atom survival probabilities for varying durations of cooling light extinction. Error bars represent Wilson score intervals (\SI{95}{\percent} confidence). The trap depth corresponding to each optical power in the trap is displayed in mK.} 
\label{SurvivalProbs} 
\end{figure}

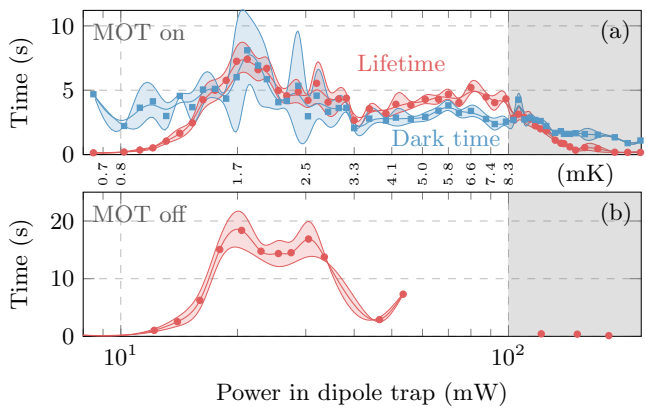
\begin{figure}[h]
	\centering
	\begin{tikzpicture}
	
	\begin{groupplot}[group style={group size=1 by 2, vertical sep=0.5cm},xmin=0,ymin=0,height=3.5cm,width=9cm]
	\nextgroupplot
	[
	ylabel={Time (s)},
	xmin=8, xmax=220,
	ymin=0, ymax=11.2,
	xticklabels={,,},
	ymajorgrids=true,
	xmajorgrids=true,
	grid style=dashed,
	xmode=log,
	clip=false
	]
	
	\addplot[
	mark size=1.2pt,
	only marks,
	mark = *,
	name path=dtdatapoints,
	color=rainbow4of8,
	/pgfplots/error bars/.cd,
	y dir = both,
	y explicit,
	] table [col sep = comma, x index = 0, y index = 1]{finalLT.csv};
	
	\addplot[
	skip coords between index={0}{39},
	color=rainbow4of8,
	/pgfplots/error bars/.cd,
	y dir = both,
	y explicit,
	] table [col sep = comma, x index = 0, y index = 1]{finalLTfilteredspline.csv};
	
	\addplot[
	style={ultra thin},
	skip coords between index={0}{39},
	name path=asdf3,
	color=rainbow4of8,
	/pgfplots/error bars/.cd,
	y dir = both,
	y explicit,
	] table [col sep = comma, x index = 0, y index = 2]{finalLTspline.csv};
	
	\addplot[
	style={ultra thin},
	skip coords between index={0}{39},
	name path=asdf4,
	color=rainbow4of8,
	/pgfplots/error bars/.cd,
	y dir = both,
	y explicit,
	] table [col sep = comma, x index = 0, y index = 3]{finalLTspline.csv};
	
	\addplot[rainbow4of8, opacity=0.2] fill between[
	of = asdf3 and asdf4,
	]; 
	
	\addplot[
	only marks,
	mark = square*,
	mark size=1pt,
	name path=dtdatapoints,
	color=rainbow1of8,
	/pgfplots/error bars/.cd,
	y dir = both,
	y explicit,
	] table [col sep = comma, x index = 0, y index = 1]{finalDT.csv};

	\addplot[
	skip coords between index={0}{39},
	name path=originaldt,
	color=rainbow1of8,
	/pgfplots/error bars/.cd,
	y dir = both,
	y explicit,
	] table [col sep = comma, x index = 0, y index = 1]{finalDTfilteredspline.csv};
	
	\addplot[
	style={ultra thin},
	skip coords between index={0}{39},
	name path=asdf1,
	color=rainbow1of8,
	/pgfplots/error bars/.cd,
	y dir = both,
	y explicit,
	] table [col sep = comma, x index = 0, y index = 2]{finalDTspline.csv};
	
	\addplot[
	style={ultra thin},
	skip coords between index={0}{39},
	name path=asdf2,
	color=rainbow1of8,
	/pgfplots/error bars/.cd,
	y dir = both,
	y explicit,
	] table [col sep = comma, x index = 0, y index = 3]{finalDTspline.csv};
	
	\addplot[rainbow1of8, opacity=0.2] fill between[
	of = asdf1 and asdf2,
	]; 
	
	\fill [fill=black!60,opacity=.2] (100,0) -- (100,11.2) -- (220,11.2) -- (220,0) -- cycle;
	
	\node[font=\small, anchor=center, color=markgrijs] at (rel axis cs:0.1,0.85) {MOT on};
	\node[anchor=west] at (rel axis cs:0.91,0.85) {\small (a)};
	
	\node[font=\small, anchor=center, color=rainbow1of8] at (rel axis cs:0.65,0.12) {Dark time};
	
	\node[font=\small, anchor=center, color=rainbow4of8] at (rel axis cs:0.57,0.64) {Lifetime};
	
	\node[anchor=center, color=black, rotate=90] at (axis cs: 9,-1.5) {\tiny 0.7};
	\node[anchor=center, color=black, rotate=90] at (axis cs: 10,-1.5) {\tiny 0.8};
	\node[anchor=center, color=black, rotate=90] at (axis cs: 20,-1.5) {\tiny 1.7};
	\node[anchor=center, color=black, rotate=90] at (axis cs: 30,-1.5) {\tiny 2.5};
	\node[anchor=center, color=black, rotate=90] at (axis cs: 40,-1.5) {\tiny 3.3};
	\node[anchor=center, color=black, rotate=90] at (axis cs: 50,-1.5) {\tiny 4.1};
	\node[anchor=center, color=black, rotate=90] at (axis cs: 60,-1.5) {\tiny 5.0};
	\node[anchor=center, color=black, rotate=90] at (axis cs: 70,-1.5) {\tiny 5.8};
	\node[anchor=center, color=black, rotate=90] at (axis cs: 80,-1.5) {\tiny 6.6};
	\node[anchor=center, color=black, rotate=90] at (axis cs: 90,-1.5) {\tiny 7.4};
	\node[anchor=center, color=black, rotate=90] at (axis cs: 100,-1.5) {\tiny 8.3};
	
	\node[anchor=center, color=black] at (axis description cs: 0.9,-0.145) {(mK)};

	\nextgroupplot
	[
	ylabel={Time (s)},
	xlabel={Power in dipole trap (mW)},
	xmin=8, xmax=220,
	ymin=0, ymax=25,
	ymajorgrids=true,
	xmajorgrids=true,
	grid style=dashed,
	xmode=log,
	]
	
	\addplot[
	mark size=1.2pt,
	only marks,
	mark = *,
	name path=dtdatapoints,
	color=rainbow4of8,
	/pgfplots/error bars/.cd,
	y dir = both,
	y explicit,
	] table [col sep = comma, x index = 0, y index = 1]{finalLTdark.csv};
	
	\addplot[
	skip coords between index={220}{1000},
	color=rainbow4of8,
	/pgfplots/error bars/.cd,
	y dir = both,
	y explicit,
	] table [col sep = comma, x index = 0, y index = 1]{finalLTdarkspline.csv};
	
	\addplot[
	style={ultra thin},
	skip coords between index={220}{1000},
	name path=asdf3,
	color=rainbow4of8,
	/pgfplots/error bars/.cd,
	y dir = both,
	y explicit,
	] table [col sep = comma, x index = 0, y index = 2]{finalLTdarkspline.csv};
	
	\addplot[
	style={ultra thin},
	skip coords between index={220}{1000},
	name path=asdf4,
	color=rainbow4of8,
	/pgfplots/error bars/.cd,
	y dir = both,
	y explicit,
	] table [col sep = comma, x index = 0, y index = 3]{finalLTdarkspline.csv};
	
	\addplot[rainbow4of8, opacity=0.2] fill between[
	of = asdf3 and asdf4,
	]; 
	
	\fill [fill=black!60,opacity=.2] (100,0) -- (100,25) -- (220,25) -- (220,0) -- cycle;
	
	\node[font=\small, anchor=center, color=markgrijs] at (rel axis cs:0.1,0.85) {MOT off};
	\node[anchor=west] at (rel axis cs:0.91,0.85) {\small (b)};
	
	\end{groupplot}
	
	\end{tikzpicture}
	\caption[An example of a floating figure]{(a) Extracted trap lifetime and dark time (unoccupied trap) as a function of trap depth under continuous illumination with \SI{780}{\nano\meter} cooling beams. (b) The extracted trap lifetimes as a function of power in the case where all cooling beams are extinguished as soon as an atom enters the trap. Solid lines indicate moving averages and the shaded regions around the data points represent single standard deviation errors extracted from exponential fits. The trap depth corresponding to each optical power in the trap is displayed in mK.}
	\label{lifetimes} 
\end{figure}

To evaluate the average occupation time of the dipole traps for a variety of trap depths, we measure the fluorescence originating from one trapping site (\autoref{Fluorfigure1}(a)) and subtract the average background counts taken from pixels that surround the traps. This compensates to a large extent any fluctuations in the background level of light at \SI{780}{\nano\meter}, which originates from atoms in the dilute MOT cloud that surrounds the dipole traps.
A histogram of the fluorescence levels is characterised by a distinct single-atom occupation peak \autoref{Fluorfigure1}(b), which provides a reliable threshold for detecting atoms. Note that higher atom occupation numbers ($N\ge2$) are not observed. Using the threshold, the lifetimes of all atoms entering the trap can be extracted and are shown to be exponentially distributed, as expected for the time between two events of a Poissonian loading process (\autoref{Fluorfigure1}(c)). To suppress the effect of the two-body loss rate on the lifetime, we perform an experiment in which all MOT beams are switched off for a fixed amount of time $\Delta t$ upon the detection of an atom. In this experiment, the survival of the atom (detection upon re-illumination) is considered a success and triggers an immediate restart of the sequence. Fluorescence traces for three different values of $\Delta t$ are shown in \autoref{Fluorfigure1}(d-f). Based on the total number of trials and successes, we calculate the survival probabilities for a range of trap depths (\autoref{SurvivalProbs}). The displayed \SI{95}{\percent} confidence interval is based on the Wilson score interval \cite{wilson1927}, which adjusts for cases with an extreme success rate or a small sample size.
Where sufficient data is available, exponential distributions fitted to these survival rates as a function time yield the average trap lifetimes. A summary of the extracted trap lifetimes is shown in \autoref{lifetimes}(a) for the case of continuous illumination with \SI{780}{\nano\meter} cooling light and in its absence (\autoref{lifetimes}(b)).
Note that suppressing the 2-body loss rate by interrupting the cooling process yields a near-threefold increase in the trap lifetime. In addition to the lifetime, the average time the trap is unoccupied (dark time) as a function of trap depth is shown in \autoref{lifetimes}(a). In the intermediate depth regime, both curves overlap, in agreement with the expected time averaged occupation probability of $0.5$. Note the discrepancy between the lifetime and the dark time for shallower traps caused by a loading inefficiency. In this region, the potential depth is small with respect to the average kinetic energy of the atom. Atoms with low initial energy may be only weakly confined and can escape more easily before reaching the centre as a consequence of background collisions or fluctuations in the laser intensity. If an atom reaches the centre of the trap and scatters sufficiently many photons to be detected, the measured lifetime is short ($<\SI{0.5}{\second}$). The importance of the dark time will become clear throughout the following section, where we discuss the objective of maximising the time averaged occupation probability (filling fraction).

\section{Maximisation of filling fraction}
\begin{figure}[t]
	\centering
	\begin{tikzpicture} 
	\begin{groupplot}[group style={group size=1 by 3,horizontal sep=0.65cm, vertical sep=0.5cm},xmin=0,ymin=0,width=4cm]
	
	\nextgroupplot
	[
	enlargelimits=false,
	axis on top,
	xmin=0, xmax=1,
	ymin=0, ymax=1,
	width=0.45\textwidth,
	height=0.1\textwidth,
	ymajorticks=false,
	xtick={0,0.1,0.2,0.3,0.4,0.5,0.6,0.7,0.8,0.9,1},
	xticklabels={0,0.1,0.2,0.3,0.4,0.5,0.6,0.7,0.8,0.9,1},
	yticklabels={,,},
	xlabel=Estimated filling fraction $\eta$,
	x label style={
		at={(axis description cs:0.5,1)},
		anchor=south,
	}
	]
	
	\addplot graphics [xmin=0,xmax=1,ymin=0,ymax=1] {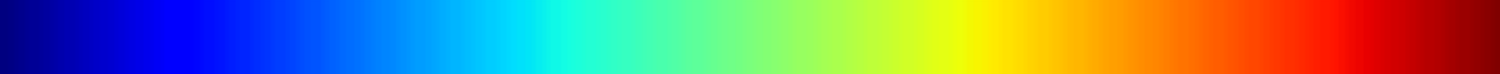};
	
	\nextgroupplot
	[
	ylabel={Holding depth (mW)},
	enlargelimits=false,
	axis on top, axis equal image,
	xmin=8, xmax=100,
	ymin=8, ymax=100,
	width=0.45\textwidth,
	height=0.5\textwidth,
	xticklabels={,,},
	ymajorgrids=true,
	xmajorgrids=true,
	yminorgrids=true,
	xminorgrids=true,
	xmode=log,
	ymode=log,
	every x tick/.style={black},
	every y tick/.style={black},
	y label style={at={(axis description cs:-0.1,.5)}},
	clip=false
	]
	
	\addplot graphics [xmin=8,xmax=100,ymin=8,ymax=100] {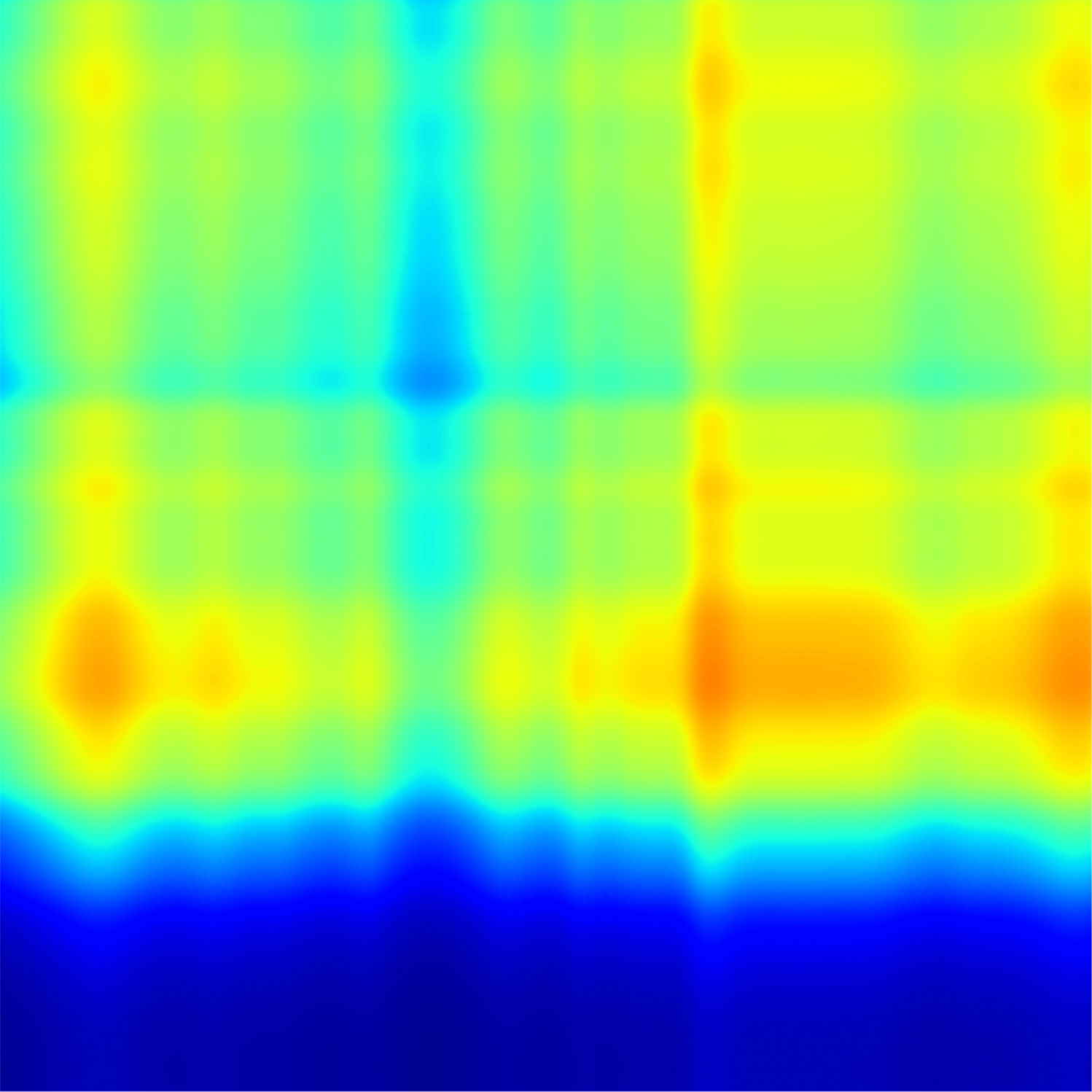};
	\draw[markrood, thick, densely dotted] (axis cs:8,47) -- (axis cs:100,47);
	\node[anchor=center, color=white] at (axis cs: 13.5,11) {\small MOT on};
	
	\node[anchor=center, color=black] at (axis cs: 7.3,20) {\tiny 20};
	\node[anchor=center, color=black] at (axis cs: 7.3,30) {\tiny 30};
	\node[anchor=center, color=black] at (axis cs: 7.3,40) {\tiny 40};
	\node[anchor=center, color=black] at (axis cs: 7.3,50) {\tiny 50};
	\node[anchor=center, color=black] at (axis cs: 7.3,60) {\tiny 60};
	\node[anchor=center, color=black] at (axis cs: 7.3,70) {\tiny 70};
	\node[anchor=center, color=black] at (axis cs: 7.3,80) {\tiny 80};
	
	\node[anchor=center, color=black] at (axis cs: 111,9) {\tiny 0.7};
	\node[anchor=center, color=black] at (axis cs: 111,10) {\tiny 0.8};
	\node[anchor=center, color=black] at (axis cs: 111,20) {\tiny 1.7};
	\node[anchor=center, color=black] at (axis cs: 111,30) {\tiny 2.5};
	\node[anchor=center, color=black] at (axis cs: 111,40) {\tiny 3.3};
	\node[anchor=center, color=black] at (axis cs: 111,50) {\tiny 4.1};
	\node[anchor=center, color=black] at (axis cs: 111,60) {\tiny 5.0};
	\node[anchor=center, color=black] at (axis cs: 111,70) {\tiny 5.8};
	\node[anchor=center, color=black] at (axis cs: 111,80) {\tiny 6.6};
	\node[anchor=center, color=black] at (axis cs: 111,90) {\tiny 7.4};
	\node[anchor=center, color=black] at (axis cs: 111,100) {\tiny 8.3};
	
	\node[anchor=center, color=black, rotate=90] at (axis description cs: 1.13,0.5) {Holding depth (mK)};
	
	\nextgroupplot
	[
	xlabel={Loading depth (mW)},
	ylabel={Holding depth (mW)},
	enlargelimits=false,
	axis on top, axis equal image,
	xmin=8, xmax=100,
	ymin=8, ymax=47,
	width=0.450\textwidth,
	height=0.50\textwidth,
	ytick={10,100},
	minor ytick={9,20,30,40,50},
	ymajorgrids=true,
	xmajorgrids=true,
	yminorgrids=true,
	xminorgrids=true,
	xmode=log,
	ymode=log,
	every x tick/.style={black},
	every y tick/.style={black},
	y label style={at={(axis description cs:-0.1,.5)}},
	clip=false
	]
	\addplot graphics [xmin=8,xmax=100,ymin=8,ymax=47] {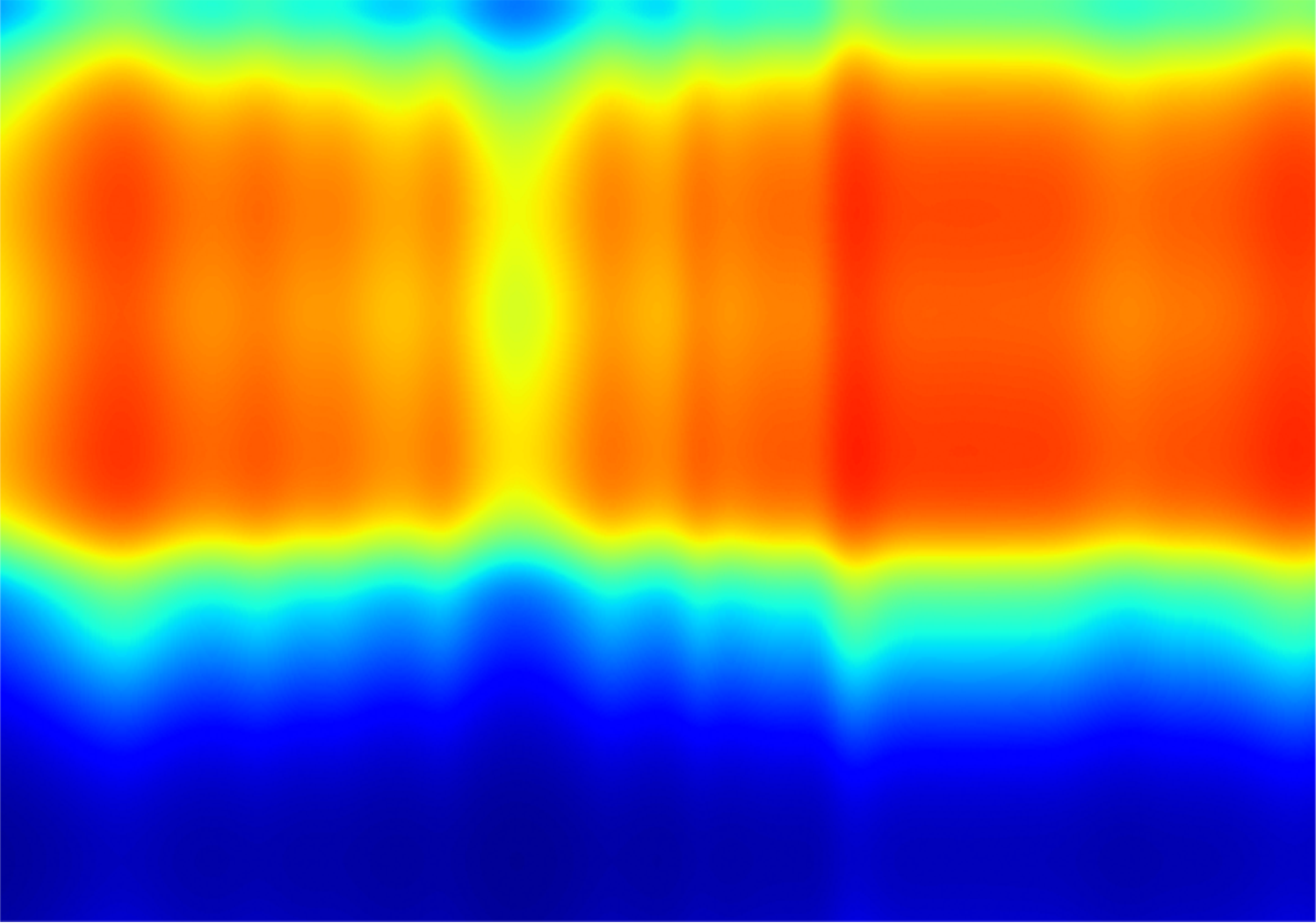};
	\draw[white] (axis cs:9,9) ;
	\node[anchor=center, color=white] at (axis cs: 13.5,11) {\small MOT off};
	
	\node[anchor=center, color=black] at (axis cs: 20,7.5) {\tiny 20};
	\node[anchor=center, color=black] at (axis cs: 30,7.5) {\tiny 30};
	\node[anchor=center, color=black] at (axis cs: 40,7.5) {\tiny 40};
	\node[anchor=center, color=black] at (axis cs: 50,7.5) {\tiny 50};
	\node[anchor=center, color=black] at (axis cs: 60,7.5) {\tiny 60};
	\node[anchor=center, color=black] at (axis cs: 70,7.5) {\tiny 70};
	\node[anchor=center, color=black] at (axis cs: 80,7.5) {\tiny 80};
	
	\node[anchor=center, color=black] at (axis cs: 7.3,20) {\tiny 20};
	\node[anchor=center, color=black] at (axis cs: 7.3,30) {\tiny 30};
	\node[anchor=center, color=black] at (axis cs: 7.3,40) {\tiny 40};
	
	\node[anchor=center, color=black] at (axis cs: 111,9) {\tiny 0.7};
	\node[anchor=center, color=black] at (axis cs: 111,10) {\tiny 0.8};
	\node[anchor=center, color=black] at (axis cs: 111,20) {\tiny 1.7};
	\node[anchor=center, color=black] at (axis cs: 111,30) {\tiny 2.5};
	\node[anchor=center, color=black] at (axis cs: 111,40) {\tiny 3.3};
	
	\node[anchor=center, color=black, rotate=90] at (axis cs: 9,51.6) {\tiny 0.7};
	\node[anchor=center, color=black, rotate=90] at (axis cs: 10,51.6) {\tiny 0.8};
	\node[anchor=center, color=black, rotate=90] at (axis cs: 20,51.6) {\tiny 1.7};
	
	\node[anchor=center, color=black, rotate=90] at (axis cs: 40,51.6) {\tiny 3.3};
	\node[anchor=center, color=black, rotate=90] at (axis cs: 50,51.6) {\tiny 4.1};
	\node[anchor=center, color=black, rotate=90] at (axis cs: 60,51.6) {\tiny 5.0};
	\node[anchor=center, color=black, rotate=90] at (axis cs: 70,51.6) {\tiny 5.8};
	\node[anchor=center, color=black, rotate=90] at (axis cs: 80,51.6) {\tiny 6.6};
	\node[anchor=center, color=black, rotate=90] at (axis cs: 90,51.6) {\tiny 7.4};
	
	\node[anchor=center, color=black, rotate=90] at (axis description cs: 1.13,0.5) {Holding depth (mK)};
	\node[anchor=center, color=black] at (axis description cs: 0.5,1.05) {(mK)};
	
	\end{groupplot}
	\end{tikzpicture}
	\caption[An example of a floating figure]{Estimated filling fractions for a feedback mechanism that switches a dipole trap from a loading to a holding depth upon detection of an atom. Estimates are calculated from measured trap dark times (loading time) and average lifetimes (holding time) in the presence (upper) and absence (lower) of cooling light after an atom has entered the trap.} 
	\label{masterplot} 
\end{figure}
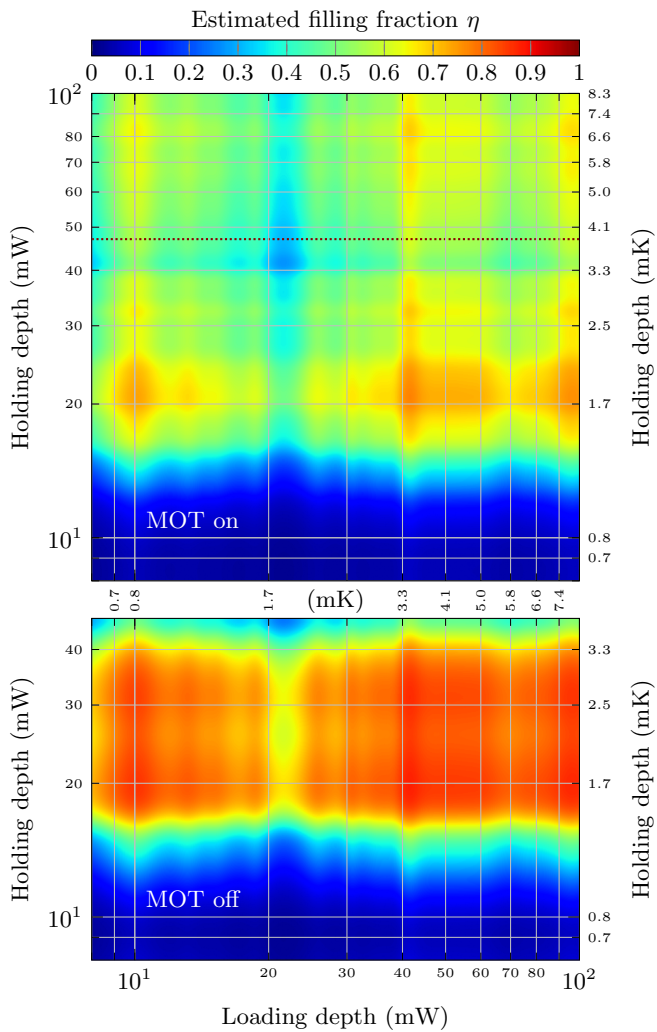

In order to achieve maximal trap occupation and beat the time averaged occupation probability of $0.5$ in the collisional blockade regime, we propose a feedback mechanism that alters the depth of a trap upon the detection of an atom. This method exploits the fast loading rate found in shallow traps (see \autoref{lifetimes}) to minimise the wait for a loading event. Once an atom is captured, the trap depth is modified, thereby prolonging its lifetime. The achievable filling fraction $\eta$ is given by the ratio
\begin{equation}
\eta = \frac{\tau}{\tau_\mathrm{d} + \tau},
\end{equation}
in which $\tau_\mathrm{d}$ and $\tau$ represent the dark time and lifetime of the trap, respectively. Since $\eta$ does not depend on the filling fraction of the trap in either regime, a high efficiency can still be achieved despite traps exhibiting filling fractions significantly below $0.5$ in the steady state. \autoref{masterplot} shows the estimated filling fraction $\eta$ for a variety of loading and holding depths of the trap.
For the calculation, we use lifetimes measured in the presence and absence of cooling beams. In the former case, two local maxima are found for a loading depth of \SI{0.8}{\milli\kelvin} at \SI{10}{\milli\watt} optical power ($\eta=0.74$) and of \SI{3.4}{\milli\kelvin} at \SI{41}{\milli\watt} optical power ($\eta=0.77$), both for a holding depth around \SI{1.7}{\milli\kelvin} at \SI{21}{\milli\watt} optical power. It should be noted that for deeper traps, captured atoms experience cooling light that is further detuned from resonance due to the increased Stark shift. For a trap depth of \SI{8.3}{\milli\kelvin} at about \SI{100}{\milli\watt} of optical power, the peaks in the bimodal distribution of the fluorescence (\autoref{Fluorfigure1}) start overlapping. This reduced signal-to-noise ratio results in an enhanced rate of false detections of atoms entering or leaving the trap. As a consequence, the extracted dark time and lifetime are significantly shortened for deep traps (shaded area in \autoref{lifetimes}(a)). This data evaluation artefact limits the practical trap depth to \SI{8.3}{\milli\kelvin}. Returning to \autoref{masterplot}, we observe that the estimated filling fractions in the absence of cooling light are found to exceed $0.8$ over an extended range of loading and holding depths, with local maxima of $\eta=0.85$ and $\eta=0.88$ at positions equal to those for continuous cooling.

\begin{figure}[t]
	\centering
	\begin{tikzpicture}
	
	\begin{groupplot}[group style={group size=2 by 1,horizontal sep=1.2cm, vertical sep=1.5cm},xmin=0,ymin=0,width=4.5cm, height=4.5cm]
	
	\nextgroupplot
	[
	xlabel={Time relative to switch (s)},
	ylabel={Occupation probability},
	x label style={at={(axis description cs:1.2, -0.3)},anchor=base},
	xmin=-10, xmax=30,
	ytick = {0,0.5,1},
	yticklabels={0,,1},
	ymin=-0.1, ymax=1.1,
	ymajorgrids=true,
	xmajorgrids=true,
	grid style=dashed,
	]
	
	\addplot[
	only marks,
	mark = *,
	mark size=0.5pt,
	color=markgrijs,
	opacity=0.5,
	/pgfplots/error bars/.cd,
	y dir = both,
	y explicit,
	] table [col sep = comma, x index = 0, y index = 1, y error index = 2]{exponentialdecay_from_trap_no_1.csv};
	
	\addplot[
	color=rainbow4of8,
	] table [col sep = comma, x index = 0, y index = 1]{exponentialdecay_from_trap_1_fit_deep.csv};
	
	\addplot[
	color=rainbow4of8,
	] table [col sep = comma, x index = 0, y index = 1]{exponentialdecay_from_trap_1_fit_shallow.csv};
	
	\node[align=left, color=rainbow4of8] at (3.5cm,1.7cm) {\small ${\tau_1}=2.0\,$s \\ \small ${\tau_2}=7.9\,$s};
	\node[anchor=west] at (rel axis cs:0.78,0.91) {\small (b)};
	
	\nextgroupplot
	[
	ymode=log,
	ymin=1e-2, ymax=1e0,
	xmin=-10, xmax=30,
	legend pos=north west,
	ymajorgrids=true,
	xmajorgrids=true,
	grid style=dashed,
	]
	
	\addplot[
	only marks,
	mark = *,
	mark size=0.5pt,
	color=markgrijs,
	opacity=0.5,
	/pgfplots/error bars/.cd,
	y dir = both,
	y explicit,
	] table [col sep = comma, x index = 0, y index = 1, y error index = 2]{exponentialdecay_from_all.csv};
	
	\addplot[
	color=markrood,
	] table [col sep = comma, x index = 0, y index = 1]{exponentialdecay_from_all_fit_deep.csv};
	
	\addplot[
	color=markrood,
	] table [col sep = comma, x index = 0, y index = 1]{exponentialdecay_from_all_fit_shallow.csv};
	
	\node[align=left, color=rainbow4of8] at (axis cs: 5, 0.02) {\small ${\tau_1}=2.0\,$s \\ \small ${\tau_2}=6.7\,$s};
	\node[anchor=west] at (rel axis cs:0.78,0.91) {\small (c)};
	
	\end{groupplot}
	
	\begin{axis}
	[
	yshift=4.3cm, 
	xlabel={Time (s)},
	ylabel={Counts (Hz)},
	xmin=0, xmax=240,
	ymin=-0.5e5, ymax=2.5e5,
	xtick={0,30,60,90,120,150,180,210,240},
	ymajorgrids=false,
	xmajorgrids=true,
	grid style=dashed,
	width=8.62cm, 
	height=3cm,
	no markers,
	]
	
	\addplot[
	color=rainbow4of8,
	] table [col sep = comma, x index = 0, y expr=\thisrowno{1}*1.3062923961932331/0.250]{trace.csv};
	
	\node[anchor=west] at (rel axis cs:0.91,0.82) {\small (a)};
	
	\node[anchor=center, color=rainbow4of8] at (axis cs:15,2.1e5) {\small S};
	\node[anchor=center, color=rainbow4of8] at (axis cs:45,2.1e5) {\small D};
	\node[anchor=center, color=rainbow4of8] at (axis cs:75,2.1e5) {\small S};
	\node[anchor=center, color=rainbow4of8] at (axis cs:105,2.1e5) {\small D};
	\node[anchor=center, color=rainbow4of8] at (axis cs:135,2.1e5) {\small S};
	\node[anchor=center, color=rainbow4of8] at (axis cs:165,2.1e5) {\small D};
	\node[anchor=center, color=rainbow4of8] at (axis cs:195,2.1e5) {\small S};
	
	\end{axis}
	
	\end{tikzpicture}
	\caption[An example of a floating figure]{(a) Fluorescence trace for a periodic (\SI{30}{\second}) modulation of the trap depth, alternating between shallow (S) and deep (D) regimes. The trap occupation probability relative to the time of switching and conditional on an atom being present at $t=0$ is shown on a linear scale for a single trap (b) and on a logarithmic scale for an ensemble average of $8$ distinct traps (c), both fitted with corresponding exponential curves. Error bars represent Wilson score intervals (b) and a single standard deviation of the ensemble (c), respectively.} 
	\label{exponential} 
\end{figure}
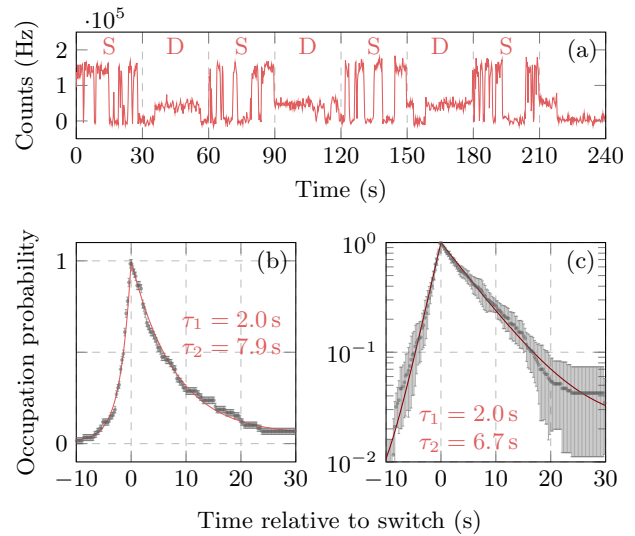

To demonstrate the concept of maximising the filling fraction of traps experimentally by changing their depth, we alter the trap depth every \SI{30}{\second} with an AOM, alternating the beam power between \SI{10}{\milli \watt} and \SI{21}{\milli\watt}. We consider only those \SI{60}{\second} long time intervals centred around the switching from shallow to deep that have an atom present upon switching. A typical fluorescence trace (\autoref{exponential}(a)) exhibits two distinct count rates for single-atom occupation that correspond to different Stark shifts associated with the trap depths.
We collect $120$ minutes of fluorescence data with a time resolution of \SI{250}{\ms} for $8$ independent traps to enhance the data acquisition rate. For each trap, we calculate the probability that an atom present at the instant of switching also occupied the trap at a particular time before or after the switch. Treating each point in time as a trial with a binary outcome (atom or no atom present), we then extract the occupation probabilities and the corresponding \SI{95}{\percent} Wilson confidence intervals. The results for one trap are shown in  \autoref{exponential}(b). The two exponentials feature different time constants that correspond to trap lifetimes of \SI{1.99\pm0.02}{\second} and \SI{7.92\pm0.15}{\second} for the shallow and deep regimes, respectively. The fact that the steady-state filling fraction of $0.49\pm0.03$ is sufficiently close to $0.5$ indicates that the lifetime in the shallow regime is very close to the average wait time $\tau_\mathrm{d}$ for a loading event. We obtain $\tau_d=\SI{2.07\pm 0.25}{\second}$, which yields a maximum achievable filling fraction of $\eta=0.79 \pm 0.02$. This is close to our previous estimate of $\eta=0.74\pm0.03$ (see \autoref{masterplot}). In order to quantify the effect of trap non-uniformity across an array of trapping sites, we calculate the ensemble average and standard deviation of the occupation probabilities for 8 independent traps. These values are shown as a function of time relative to the switch in \autoref{exponential}(c). The exponential time constants are \SI{2.00\pm0.07}{\second} and \SI{6.67\pm0.08}{\second}, the latter of which differs substantially from the individual trap result. This discrepancy is likely due to the local variation in trap depth and shape, as well as their position with respect to the centre of the MOT.

\section{Conclusion}
In conclusion, we have demonstrated that through modulation of the trap depth it is possible to push the average occupation probability of a dipole-force trap in the collisional blockade regime beyond the limited value of $0.5$. This method remains effective in the case of steady-state filling fractions below $0.5$, as long as there is a sufficiently large difference between the dark time and lifetime of the loading and holding regimes, respectively. Variations of the scheme involve either increasing or decreasing the trap from its loading depth upon the detection of an atom. Our proof of concept can be applied to simultaneous maximisation of filling fractions in tweezer arrays through the use of occupation-triggered holograms capable of modulating individual trapping sites on the fly. This scheme relies on an EMCCD camera continuously monitoring each trapping site, followed by an update of the hologram that increases the depth of those traps that acquired a single atom. Standard algorithms can be used to pre-compute such holograms for all required depth combinations. This scheme involves a number of operations equal to the size of the array to be filled, providing an efficient alternative to traditional filling schemes that rely on complex relocation operations.

This project received funding from the European Union’s Horizon 2020 research and innovation programme under the Marie
Sklodowska Curie grant agreement No. 765075 and EPSRC through the quantum technologies program (NQIT hub, No. EP/M013243/1).

\bibliography{main}

\end{document}